\documentclass[aps,pra,twocolumn]{revtex4-2}
\usepackage{graphicx}
\usepackage{xcolor}
\usepackage{mathtools}
\usepackage{comment}
\usepackage{braket}
\usepackage{ulem}
\usepackage{threeparttable}

\begin{document}

\title{Nonlocal correlations in quantum energy teleportation: perspectives from their Majorana representations and information thermodynamics}

\author{Hiroaki Matsueda\thanks{hiroaki.matsueda.c8@tohoku.ac.jp}${}^{1,2}$, Yusuke Masaki${}^{1}$, Kanji Itoh${}^{1}$, Atsushi Ono${}^{3}$, and Joji Nasu${}^{3}$}

\affiliation{
${}^{1}$Department of Applied Physics, Graduate School of Engineering, Tohoku University, Sendai 980-8579, Japan \\
${}^{2}$Center for Science and Innovation in Spintronics, Tohoku University, Sendai 980-8577, Japan \\
${}^{3}$Department of Physics, Graduate School of Science, Tohoku University, Sendai 980-8578, Japan
}
\date{\today}
\begin{abstract}
Motivated by anomalous nonlocal correlation in the Kitaev spin liquids, we propose a quantum energy teleportation protocol between remote partners Alice and Bob on a quantum spin model, and examine how its performance is characterized by Majorana fermions that clearly depict nonlocal correlations inherent in the model. In our model, Bob's energy extraction is activated by local energy injection by Alice's projective measurement and subsequent classical communication of the measurement result. We derive two formulae: one for the maximally extracted energy by the protocol and the other for the maximum of energy reduction at Bob's local site. We find that the extracted energy becomes positive when a nonlocal correlator defined by Majorana fermions at Alice's and Bob's sites is finite. We also find that the amount of the energy reduction becomes positive when another nonlocal Majorana correlator is finite. In both formulae, the correlators appear as a result of Bob's feedback unitary operation. We discuss effective information-thermodynamical aspects behind the protocol at zero temperature.
\end{abstract}

\maketitle

\section{Introduction}
\label{introduction}

Quantum teleportation transfers quantum information (or a quantum state) to a distant place by using an entanglement pair and projective measurement. Quantum teleportation is thus the most fundamental protocol for quantum communication technologies, and nearly 30 years have passed since this protocol was developed~\cite{Bennett}. Quantum teleportation provides meaningful perspectives not only in the field of communication technologies, but also in recent researchs on traversable wormholes in spacetime physics and its holographic realization on quantum processors~\cite{Susskind,Schuster,Jafferis,Popov}.

Phenomena similar to quantum teleportation can also be seen in solid-state physics, although the transportation range seems to be of microscopic scale. Nice playgrounds are Kitaev quantum spin systems. The Kitaev model is an exactly solvable quantum spin model, whose ground state is a quantum spin liquid with elementary excitations described by Majorana fermions~\cite{Kitaev,Nussinov,Hermanns,Knolle,Takagi,Janssen,Motome,Hickey}. Two related studies have been conducted in recent years. The first one is about detection of the magnetic moment at one edge of the Kitaev model after pulse magnetic excitation to the opposite edge~\cite{Minakawa,Koga,Taguchi,Nasu,Misawa}. The second one is about long-range correlation between two remote impulities introduced in the Kitaev model~\cite{Takahashi}. In both cases, there is no direct transportation of spin waves. Instead, the Majorana physics is a key for understanding such long-range correlations. In addition, performing braiding operations of Majorana zero modes is a candidate for teleportation-based quantum information processing~\cite{Vijay,Fu,Huang}. Thus, the physics of Majorana fermions has good connection with quantum teleportation theories and technologies.

Unfortunately, the abovementioned two studies do not directly consider how each physical process and the long-range correlation correspond to traditional quantum teleportation protocols. In the protocol, any nonlocal correlation function between an information sender and receiver does not appear. In order to understand the roles of the Majorana correlations on the teleportation-like behaviors in the Kitaev systems, we consider quantum energy teleportation (QET) rather than quantum teleportation~\cite{Hotta,Hotta2,Hotta3,Hotta4,Trevison,Rodriguez,Ikeda,Ikeda2,Itoh,Wang}. As shown later, this research is also related to a very fundamental question about the second law of information thermodynamics, which is how work extraction is enhanced by quantum information~\cite{Sagawa,Tajima,Funo,Park,Manzano,Minagawa,Itoh2}.

Motivated by these cross-disciplinary interests, we examine a QET protocol for remote partners Alice and Bob on the ground state of a one-dimensional quantum spin model that is transformed into a model of Majorana fermions. In a basic assumption of the QET protocol, Alice and Bob perform local operations and classical communication by a speed much faster than the transfer of any physical modes. Thus, this protocol does not demand direct physical connection between Alice and Bob except for the ground-state entanglement.
In this situation, Bob's feedback unitary transformation activates energy extraction after local energy injection by Alice's projective measurement. We would like to understand how the entanglement (or nonlocal Majorana correlation) facilitates the positive energy extraction by the QET protocol. We will derive the formula to represent the relation between the extracted energy and a nonlocal Majorana correlator. We will also derive the formula that relates another nonlocal Majorana correlator to local energy reduction at Bob's site, and find that the result is consistent with our recent work on the upper bound on locally extractable energy from an entangled pure state under feedback control~\cite{Itoh2}. These nonlocal Majorana correlators appear as a result of Bob's feedback operation, and contribute to the energy extraction and the local energy reduction. If we interpret the extractable energy as work, it is a thermodynamically natural question to ask whether 
the QET involves heat transfer between Bob's local system and the other part of the whole system. Even though we consider the QET protocol at zero temperature, it is possible to define heat as the difference between the amount of the local energy reduction and the work. The heat originates in the change in the interaction energy between the local system and the other part. We will find that the heat is finite in general and becomes zero for the QET process that maximizes the work. We will also mention the second law of effective information thermodynamics for Bob's local system in which the effective temperature originates in the entanglement between the local system and the other part of the whole system. We will find that the nonlocal Majorana correlator related to the local energy reduction is different from the quantum-classical (QC) mutual information appearing in our previous work~\cite{Itoh2}, although both of the Majorana correlator and the QC-mutual information are quantum-information resources for driving these protocols.

This paper is organized as follows. In the next section, we introduce our model, and represent it using Jordan-Wigner fermions. In section~\ref{protocol}, we explain our QET protocol, and derive the formulae of the extracted energy and the local energy reduction. In section~\ref{Majorana}, we derive the Majorana fermion representations of these formulae, and discuss the roles of the nonlocal Majorana correlations on these formulae. Section~\ref{IT} is devoted to the discussion about the first and second laws of effective information thermodynamics inherent in the present model. On the basis of the present result, we summarize our results in section~\ref{conclusion}. In Appendix~\ref{appA}, the symmetry operations to determine the ground-state properties are provided. In Appendix~\ref{appB}, mathematical details for deriving the maxima of the extracted energy and the local energy reduction are provided. In Appendix~\ref{appC}, the effect of the degeneracy of even- and odd-parity states on the QET performance is provided. The derivations of some equations are also provided in Appendices~\ref{appD} and \ref{appE}.

\section{Model}
\label{model}

The QET protocol is as follows: First, the ground state with quantum entanglement is prepared, and Alice makes a local measurement at one endpoint of the system. Next, the measurement result is transmitted to Bob, at the opposite endpoint, via a classical communication channel. Then, Bob performs a feedback unitary operation based on the result. Alice injects energy locally into the system by the measurement. Bob can extract energy from the system by the unitary operation. It is assumed that Alice's classical communication and Bob's manipulation are sufficiently faster than the transmission rate of elementary excitation that can occur after Alice's measurement.

\begin{figure}[htbp]
\begin{center}
\includegraphics[width=7.5cm]{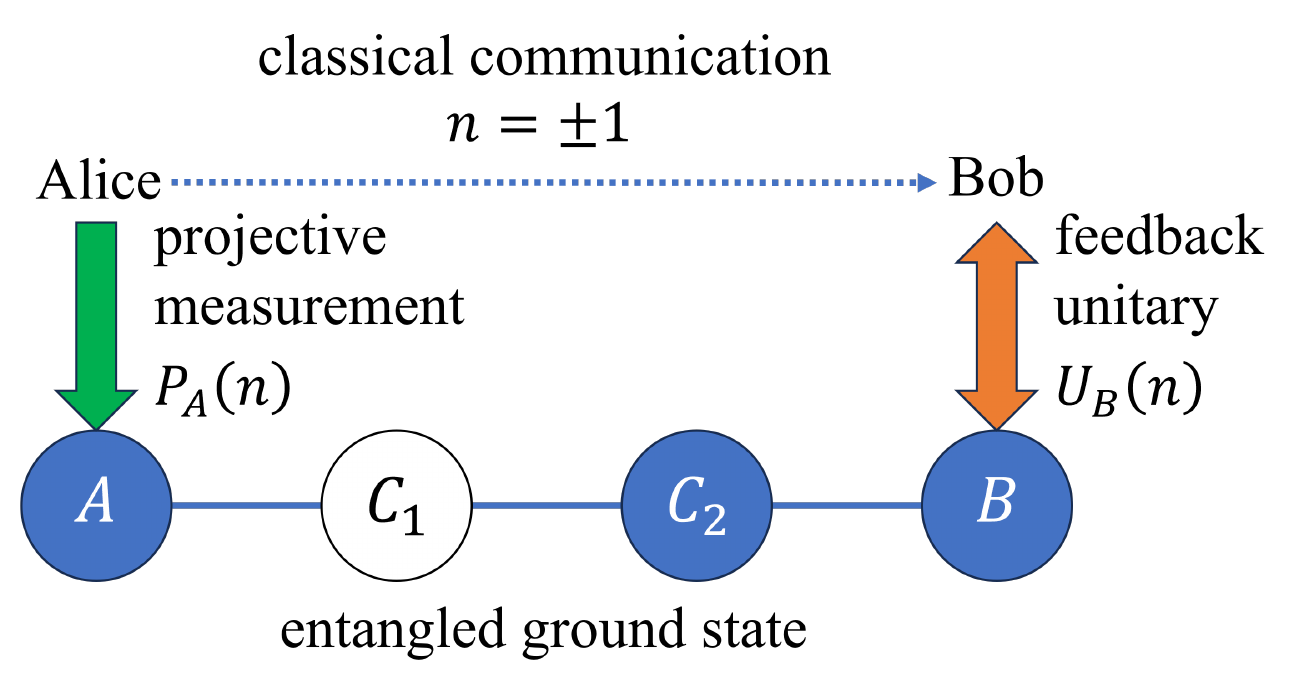}
\end{center}
\caption{
Illustration of our four-spin model and protocol. Alice and Bob perform local projective measurement $P_{A}(n)$ to spin $A$ and feedback unitary operation $U_{B}(n)$ to spin $B$, respectively. The green arrow at site $A$ represents energy injection by the measurement, and the orange arrow at site $B$ represents energy injection or extraction depending on the selection of the unitary operator. The classical communication between Alice and Bob is indicated by the dotted arrow, and the measurement results are $n=\pm 1$. The final result for the extracted energy is represented by the operators defined at sites $A$, $C_{2}$, and $B$ (blue shaded circles).
}
\label{KitaevQETfig1}
\end{figure}

Let us start with the Hamiltonian defined by
\begin{align}
H&=H_{A}+V+H_{B} , \\
H_{A}&=h\sigma_{A}^{z} , \\
V&=k\left(\sigma_{A}^{x}\sigma_{C_{1}}^{x}+\sigma_{C_{1}}^{y}\sigma_{C_{2}}^{y}+\sigma_{C_{2}}^{x}\sigma_{B}^{x}\right) , \\
H_{B}&=h\sigma_{B}^{z},
\end{align}
where $\sigma_{i}^{\alpha}$ ($\alpha=x,y,z$) is the Pauli operator at site $i$ (the site indices, $A$, $C_{1}$, $C_{2}$, and $B$, are sometimes replaced with $0$, $1$, $2$, and $3$, respectively). The model contains four sites $A$, $C_{1}$, $C_{2}$, and $B$, and their spatial geometry is illustrated in Fig.~\ref{KitaevQETfig1}. The interaction strength $k$ creates entanglement, and the edge magnetic field $h$ controls the amount of the entanglement. The interaction term has a special form so that our model matches with the analysis of Majorana correlation in the Kitaev spin liquids. This point will be discussed in Sec.~\ref{Majorana}. We will apply the local operation and classical communication to this ground state.

For later convenience to discuss Majorana correlation, we introduce the Jordan-Wigner transformation to spin operators:
\begin{align}
\sigma_{i}^{z}&=2f_{i}^{\dagger}f_{i}-1 , \\
\sigma_{i}^{x}&=\Theta_{i}\bigl(f_{i}^{\dagger}+f_{i}\bigr) , \\
\sigma_{i}^{y}&=-i\Theta_{i}\bigl(f_{i}^{\dagger}-f_{i}\bigr) , \\
\Theta_{i}&=e^{i\pi\sum_{l(<i)}f_{l}^{\dagger}f_{l}}=\prod_{l(<i)}\left(-\sigma_{l}^{z}\right) , 
\end{align}
where $i=0,1,2,3$ ($A,C_{1},C_{2},B$), $f_{i}^{\dagger}$ and $f_{i}$ are the creation and annihilation operators of the Jordan-Wigner fermion at site $i$, respectively, and $\Theta_{i}$ is the string operator ($\Theta_{0}=1$). Hereafter we abbreviate $\Theta_{3}$($=\Theta_{B}$) as $\Theta$. The interaction term is transformed into
\begin{align}
V=&k\bigl(f_{A}^{\dagger}-f_{A}\bigr)\bigl(f_{C_{1}}^{\dagger}+f_{C_{1}}\bigr) \nonumber \\
&-k\bigl(f_{C_{1}}^{\dagger}+f_{C_{1}}\bigr)\bigl(f_{C_{2}}^{\dagger}-f_{C_{2}}\bigr) \nonumber \\
&+k\bigl(f_{C_{2}}^{\dagger}-f_{C_{2}}\bigr)\bigl(f_{B}^{\dagger}+f_{B}\bigr).
\end{align}

\begin{figure}[htbp]
\begin{center}
\includegraphics[width=7cm]{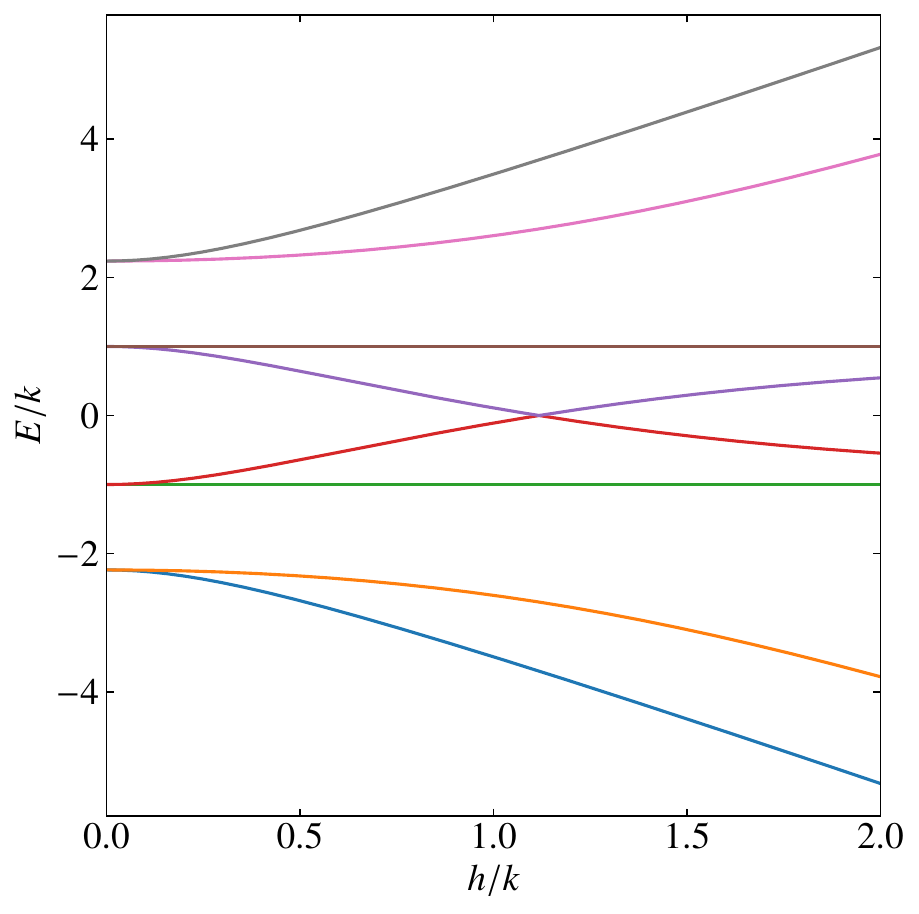}
\end{center}
\caption{
Eigenvalue spectra $E$ in the even-parity sector as a function of $h$. The lowest line represents the lowest value of $\epsilon$ in Eq.~(\ref{eigen}).
}
\label{KitaevQETfig2}
\end{figure}

Figure~\ref{KitaevQETfig2} shows the $h$ dependence of the eigenvalue spectra of $H$ in the even-parity sector that is spanned by even numbers of Jordan-Wigner fermions. The eigenvalue spectra are degenerate with those in the odd-parity sector. We find symmetry between negative and positive spectra. We also find twofold degeneracies for all of the eigenstates in each parity sector for $h=0$, and the degeneracies are lifted by introducing finite $h$ values. In Appendix~\ref{appA}, we summarize symmetry operations associated with these spectroscopic properties.

As a playground of QET, we prepare the lowest-energy eigenstate $\left|\psi\right>$ in the even-parity sector. The state $\left|\psi\right>$ for $h\ge 0$ is given by
\begin{align}
\left|\psi\right>=&Z\left[\left(\left|eeff\right>+\left|efef\right>\right)+\alpha\left(\left|ffff\right>+\left|feef\right>\right) \right. \nonumber \\
&\left.+\left(\left|ffee\right>+\left|fefe\right>\right)+\beta\left(\left|eeee\right>+\left|effe\right>\right)\right] ,
\end{align}
where $\left|eeff\right>=f_{C_{2}}^{\dagger}f_{B}^{\dagger}\left|eeee\right>$, $\left|eeee\right>=\left|e\right>_{A}\otimes\left|e\right>_{C_{1}}\otimes\left|e\right>_{C_{2}}\otimes\left|e\right>_{B}$, $\left|eeee\right>$ is the vacuum of the Jordan-Wigner fermions, and $\sigma_{i}^{z}\left|e\right>_{i}=-\left|e\right>_{i}$, $\sigma_{i}^{z}\left|f\right>_{i}=\left|f\right>_{i}$ with $i=A,C_{1},C_{2},B$. The parameters, $\alpha$ and $\beta$, are respectively defined by
\begin{align}
\alpha&=\frac{2k}{\epsilon+k-2h} \label{alp}, \\
\beta&=\frac{2k}{\epsilon+k+2h} \label{bt},
\end{align}
with the lowest energy $\epsilon=k\left(\alpha+\beta+1\right)$, and the normalization factor $Z$ is defined by
\begin{align}
\left(4+2\alpha^{2}+2\beta^{2}\right)Z^{2}=1. \label{A}
\end{align}
For $\alpha$ and $\beta$, we find the following equation:
\begin{align}
k\left(\alpha-\beta\right)=2h\alpha\beta. \label{ab}
\end{align}
Here, the lowest energy $\epsilon$ is determined by solving the following equation:
\begin{eqnarray}
\left(\epsilon+k\right)\left(\epsilon^{2}-5k^{2}\right)-4h^{2}\left(\epsilon-k\right)=0. \label{eigen}
\end{eqnarray}
We find that $\lim_{h\rightarrow 0}\epsilon=-\sqrt{5}k$ and $\epsilon< -\sqrt{5}k$ for $h>0$ as shown in Fig.~\ref{KitaevQETfig2}. By using Eqs.~(\ref{alp}), (\ref{bt}), and (\ref{eigen}), we can derive $\alpha\beta=(\epsilon-k)/(\epsilon+k)>0$ for general $h$ values. The expectation values of Hamiltonian terms $H_{A}$, $V$, and $H_{B}$ by $\left|\psi\right>$ are evaluated as
\begin{align}
\epsilon&=\epsilon_{A}+\epsilon_{V}+\epsilon_{B} , \\
\epsilon_{A}&=\left<\psi\right|H_{A}\left|\psi\right>=2hZ^{2}\left(\alpha^{2}-\beta^{2}\right) \le 0 , \\
\epsilon_{V}&=\left<\psi\right|V\left|\psi\right>=\epsilon-\epsilon_{A}-\epsilon_{B}=\epsilon-2\epsilon_{A} , \\
\epsilon_{B}&=\left<\psi\right|H_{B}\left|\psi\right>=\epsilon_{A} .
\end{align}
For later convenience, we introduce the following representation for $V$: $V=H_{L}+H_{C}+H_{R}$, $H_{L}=k\sigma_{A}^{x}\sigma_{C_{1}}^{x}$, $H_{C}=k\sigma_{C_{1}}^{y}\sigma_{C_{2}}^{y}$, and $H_{R}=k\sigma_{C_{2}}^{x}\sigma_{B}^{x}$. The expectation values of $H_{L}$ and $H_{R}$ by $\left|\psi\right>$ are given by
\begin{align}
\epsilon_{L}&=\left<\psi\right|H_{L}\left|\psi\right>=4kZ^{2}\left(\alpha+\beta\right)<0 , \\
\epsilon_{R}&=\left<\psi\right|H_{R}\left|\psi\right>=\epsilon_{L} .
\end{align}

For $h=0$, $\left|\psi\right>$ is factorized into the following form:
\begin{align}
\left|\psi\right>=&Z\left|fe-ef\right>_{AB}\otimes\left|fe+ef\right>_{C_{1}C_{2}} \nonumber \\
&+Z\alpha\left|ff+ee\right>_{AB}\otimes\left|ff+ee\right>_{C_{1}C_{2}},
\end{align}
where we have rearranged the order of the basis states and $\left|ff+ee\right>_{AB}=\left|f\right>_{A}\otimes\left|f\right>_{B}+\left|e\right>_{A}\otimes\left|e\right>_{B}$. It is clear that the local states at Alice's and Bob's sites are correlated with each other. This correlation characterizes long-range entanglement in this model at least for $h=0$.

\section{Protocol}
\label{protocol}

Let us introduce Alice's measurement and Bob's feedback unitary operations in our QET protocol as already indicated in Fig.~\ref{KitaevQETfig1}. Alice makes the following local measurements on spin $A$:
\begin{align}
&P_{A}(n)=\frac{1}{2}\left(I_{A}+n\vec{r}\cdot\vec{\sigma}_{A}\right), \\
&\vec{r}=\left(r_{x},r_{y},r_{z}\right)=\left(\sin\mu\cos\nu,\sin\mu\sin\nu,\cos\mu\right),
\end{align}
where the measurement result $n$ takes $\pm 1$, $I_{A}$ is the identity operator, $\vec{r}$ is a unit vector, and $P_{A}^{2}(n)=P_{A}(n)$. Alice sends the measurement result $n$ to Bob by the classical communication channel. Depending on the result $n$, Bob performs the following feedback unitary operation on spin $B$:
\begin{align}
&U_{B}(n)=e^{in\theta\vec{s}\cdot\vec{\sigma}_{B}}=I_{B}\cos\theta+in\vec{s}\cdot\vec{\sigma}_{B}\sin\theta, \label{UB} \\
&\vec{s}=\left(s_{x},s_{y},s_{z}\right)=\left(\sin\xi\cos\eta,\sin\xi\sin\eta,\cos\xi\right) ,
\end{align}
where $\vec{s}$ is a unit vector. For $\theta\ne0,\pm\pi/2$, the feedback unitary operation is realized.

We evaluate the energy injected into Alice's local site by Alice's measurement, $\Delta E_{A}=E_{A}-\epsilon$, where $E_{A}$ is the energy expectation value after Alice's measurement:
\begin{align}
E_{A}=\sum_{n=\pm 1}\left<\psi\right| P_{A}(n)HP_{A}(n)\left|\psi\right> . \label{EA}
\end{align}
Here, the quantum state after Alice's measurement is represented as $\left|\psi_{A}(n)\right>=P_{A}(n)\left|\psi\right>/\sqrt{\left<\psi\right|P_{A}(n)\left|\psi\right>}$, and Eq.~(\ref{EA}) is reduced to $E_{A}=\sum_{n=\pm 1}p_{n}\left<\psi_{A}(n)\right|H\left|\psi_{A}(n)\right>$, where $p_{n}=\left<\psi\right|P_{A}(n)\left|\psi\right>$ is the probability of obtaining a measurement result $n$. The energy $E_{A}$ is evaluated as
\begin{align}
E_{A}&=\sum_{n=\pm 1}\left<\psi\right|\left[P_{A}(n),H_{A}+H_{L}\right]P_{A}(n)\left|\psi\right>+\epsilon \nonumber \\
&=r_{z}^{2}\epsilon_{A}+r_{x}^{2}\epsilon_{L}+\epsilon_{C}+\epsilon_{R}+\epsilon_{B},
\end{align}
and the injected energy, $\Delta E_{A}$, is given by
\begin{align}
\Delta E_{A}=\left(r_{z}^{2}-1\right)\epsilon_{A}+\left(r_{x}^{2}-1\right)\epsilon_{L}.
\end{align}
It is clear that Alice's measurement injects positive energy ($\Delta E_{A}>0$) into the local site $A$.

We also introduce the energy expectation value after Bob's feedback unitary operation:
\begin{align}
E_{B}=\sum_{n=\pm 1}\left<\psi\right|P_{A}(n)U_{B}^{\dagger}(n)HU_{B}(n)P_{A}(n)\left|\psi\right> ,
\end{align}
which is evaluated as
\begin{align}
E_{B}=&\sum_{n=\pm 1}\left<\psi\right|\left[P_{A}(n),H_{A}+H_{L}\right]P_{A}(n)\left|\psi\right>+\epsilon_{A}+\epsilon_{L}+\epsilon_{C} \nonumber \\
&+\sum_{n=\pm 1}\left<\psi\right|P_{A}(n)U_{B}^{\dagger}(n)\left(H_{R}+H_{B}\right)U_{B}(n)\left|\psi\right> \nonumber \\
=&r_{z}^{2}\epsilon_{A}+r_{x}^{2}\epsilon_{L}+\epsilon_{C} \nonumber \\
&+\sum_{n=\pm 1}\left<\psi\right|P_{A}(n)U_{B}^{\dagger}(n)\left(H_{B}+H_{R}\right)U_{B}(n)\left|\psi\right> .
\end{align}
The extracted energy by Bob, $\Delta E_{B}=E_{A}-E_{B}$ (the energy extraction is possible for $\Delta E_{B}>0$), is given by the sum of two contributions:
\begin{align}
\Delta E_{B}=\Delta E_{B,B}+\Delta E_{B,R} , \label{BBI}
\end{align}
where
\begin{align}
\Delta E_{B,i}=\epsilon_{i}-\sum_{n=\pm 1}\left<\psi\right|P_{A}(n)U_{B}^{\dagger}(n)H_{i}U_{B}(n)\left|\psi\right> , \label{DEBi}
\end{align}
with $i=B,R$. The quantity $\Delta E_{B,B}$ corresponds to the reduction of internal energy of Bob's local system. We have confirmed that Eqs.~(\ref{BBI}) and (\ref{DEBi}) hold regardless of the details of $H_{C}$ under the constraint that $H_{C}$ does not contain the operators defined at sites $A$, $C_{2}$, and $B$, since $\epsilon_{C}$ disappears by subtracting $E_{A}$ and $E_{B}$. The second term of $\Delta E_{B,i}$ corresponds to nonlocal correlation between Alice's measurement operator $P_{A}(n)$ and the operator $U_{B}^{\dagger}(n)H_{i}U_{B}(n)$ that is associated with Bob's feedback operation. The quantities $\Delta E_{B,B}$ and $\Delta E_{B,R}$ are evaluated as follows:
\begin{align}
\Delta E_{B,B}=&\epsilon_{B}\left(1-s_{z}^{2}\right)\left(1-\cos\left(2\theta\right)\right) \nonumber \\
&-h\left(r_{x}s_{y}C_{AB}-r_{y}s_{x}D_{AB}\right)\sin\left(2\theta\right) , \label{DEBB} \\
\Delta E_{B,R}=&\epsilon_{R}\left(1-s_{x}^{2}\right)\left(1-\cos\left(2\theta\right)\right) \nonumber \\
&+kr_{x}s_{y}C_{AR}\sin\left(2\theta\right) , \label{DEBVB}
\end{align}
where the correlators, $C_{AB}$, $D_{AB}$, and $C_{AR}$ are, respectively, defined by
\begin{align}
C_{AB}&=\left<\psi\right|\sigma_{A}^{x}\sigma_{B}^{x}\left|\psi\right>=4Z^{2}\left(1+\alpha\beta\right) , \label{CAB} \\
D_{AB}&=\left<\psi\right|\sigma_{A}^{y}\sigma_{B}^{y}\left|\psi\right>=4Z^{2}\left(1-\alpha\beta\right) \label{DAB}, \\
C_{AR}&=\left<\psi\right|\sigma_{A}^{x}\sigma_{C_{2}}^{x}\sigma_{B}^{z}\left|\psi\right>=4Z^{2}\left(\alpha-\beta\right) .
\end{align}
For $h\to 0$ ($\alpha=\beta$ and $\alpha\beta=(3+\sqrt{5})/2>1$), we find $C_{AB}=1$, $D_{AB}<0$, and $C_{AR}=0$. In order to achieve $\Delta E_{B}=\Delta E_{B,B}+\Delta E_{B,R}>0$, at least either $\Delta E_{B,B}$ or $\Delta E_{B,R}$ must be positive. Since the first terms in Eqs.~(\ref{DEBB}) and (\ref{DEBVB}) are always negative, their second terms are essential to $\Delta E_{B}>0$. The second terms are finite values for the feedback operation ($\theta\ne 0,\pm\pi/2$), at which non-local correlations $C_{AB}$, $D_{AB}$, and $C_{AR}$ appear.

If we replace $U_{B}(n)$ with an unitary operation without feedback, $u_{B}$ ($=U_{B}(+1)$ or $U_{B}(-1)$), in Eq.~(\ref{DEBi}), we find
\begin{align}
&\epsilon_{B}-\sum_{n=\pm 1}\left<\psi\right|P_{A}(n)u_{B}^{\dagger}H_{B}u_{B}\left|\psi\right> \nonumber \\
&\;\;\;\; =\epsilon_{B}\left(1-s_{z}^{2}\right)\left(1-\cos\left(2\theta\right)\right) , \\
&\epsilon_{R}-\sum_{n=\pm 1}\left<\psi\right|P_{A}(n)u_{B}^{\dagger}H_{R}u_{B}\left|\psi\right> \nonumber \\
&\;\;\;\; =\epsilon_{R}\left(1-s_{x}^{2}\right)\left(1-\cos\left(2\theta\right)\right) .
\end{align}
Here, the absence of $n$ dependence of $u_{B}$ leads to cancellation of the effect of Alice's measurement $\sum_{n}P_{A}(n)=1$. The results support our statement that the feedback control induces the effects of the nonlocal correlations on $\Delta E_{B,B}$ and $\Delta E_{B,R}$.

By optimizing $\vec{r}$, $\vec{s}$, and $\theta$ (see Appendix~\ref{appB}), we obtain the maximum of $\Delta E_{B}$, denoted by $\Delta E_{B}^{\rm max}$, as
\begin{align}
\Delta E_{B}^{\rm max}&=\sqrt{\epsilon_{B}^{2}+\left(hD_{AB}\right)^{2}}-\left|\epsilon_{B}\right|, \label{DEBmax}
\end{align}
where the optimization conditions are given by
\begin{align}
&\vec{r}=(0,1,0) , \\
&\vec{s}=(1,0,0) , \\
&\sin(2\theta)=\frac{hD_{AB}}{\sqrt{\epsilon_{B}^{2}+(hD_{AB})^{2}}} , \\
&\cos(2\theta)=\frac{-\epsilon_{B}}{\sqrt{\epsilon_{B}^{2}+(hD_{AB})^{2}}} .
\end{align}
For the parameters that maximize $\Delta E_{B}$, we find that $\Delta E_{B,B}$ is equal to $\Delta E_{B}^{\rm max}$ and $\Delta E_{B,R}$ becomes zero:
\begin{align}
\Delta E_{B,B}&=\epsilon_{B}+\sqrt{\epsilon_{B}^{2}+\left(hD_{AB}\right)^{2}}=\Delta E_{B}^{\rm max} , \\
\Delta E_{B,R}&=0.
\end{align}
These results show that a part of the second term in Eq.~(\ref{DEBB}), $hr_{y}s_{x}D_{AB}\sin(2\theta)$, is crucial for the positive energy extraction.

We also derive the maximum value of $\Delta E_{B,B}$, denoted by $\Delta E_{B,B}^{\rm max}$, which is a primary focus in our previous study~\cite{Itoh2} (see Appendix~\ref{appB}). The maximum value is obtained as
\begin{align}
\Delta E_{B,B}^{\rm max}&=\sqrt{\epsilon_{B}^{2}+\left(hC_{AB}\right)^{2}}-\left|\epsilon_{B}\right|, \label{DEBBmax}
\end{align}
where the optimization conditions are given by
\begin{align}
&\vec{r}=(1,0,0) , \\
&\vec{s}=(0,1,0) , \\
&\sin(2\theta)=\frac{-hC_{AB}}{\sqrt{\epsilon_{B}^{2}+(hC_{AB})^{2}}} \label{sine} , \\
&\cos(2\theta)=\frac{-\epsilon_{B}}{\sqrt{\epsilon_{B}^{2}+(hC_{AB})^{2}}} \label{cosine} .
\end{align}
For the parameters that maximize $\Delta E_{B,B}$, we find that $\Delta E_{B,R}$ is always negative:
\begin{align}
\Delta E_{B,R}=\epsilon_{R}\left(1-\cos(2\theta)\right)+kC_{AR}\sin(2\theta) <0. \label{debin}
\end{align}
Because of the inequality $\left|C_{AB}\right|>\left|D_{AB}\right|$, $\Delta E_{B,B}^{\rm max}$ is larger than $\Delta E_{B}^{\rm max}$. However, because of the negative feature of $\Delta E_{B,R}$, the maximum of $\Delta E_{B}$ can not be reached by optimizing $\Delta E_{B,B}$.

\begin{figure}[htbp]
\begin{center}
\includegraphics[width=7cm]{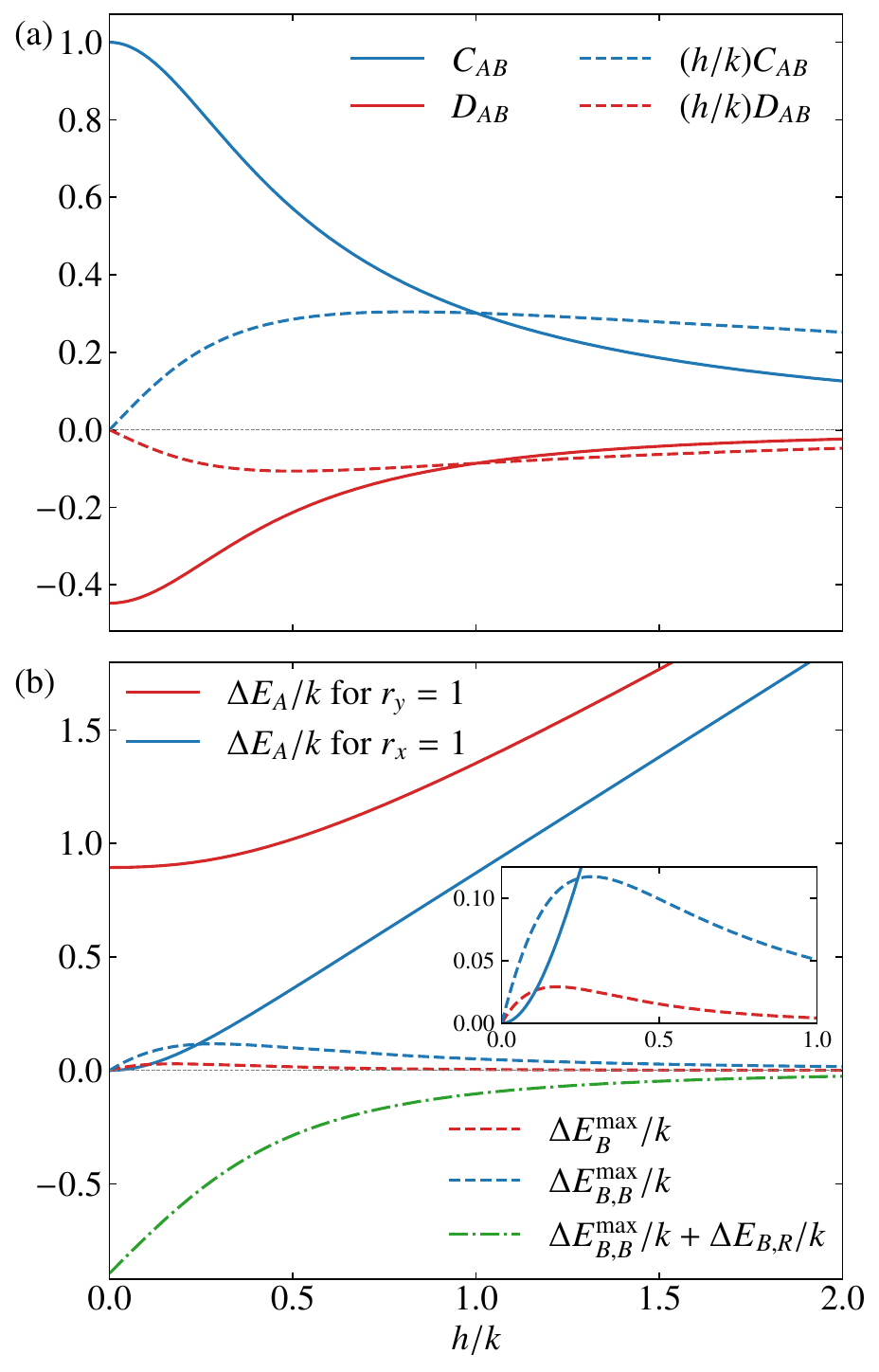}
\end{center}
\caption{(a) Nonlocal correlators $C_{AB}$ and $D_{AB}$ ($hC_{AB}$ and $hD_{AB}$) as a function of $h$. (b) $\Delta E_{A}$ for $r_{y}=1$, $\Delta E_{A}$ for $r_{x}=1$, $\Delta E_{B}^{\rm max}$, $\Delta E_{B,B}^{\rm max}$, and the sum of Eq.~(\ref{DEBBmax}) and Eq.~(\ref{debin}), as a function of $h$. The inset shows a magnified view for $h<k$.
}
\label{KitaevQETfig3}
\end{figure}

Figure~\ref{KitaevQETfig3}(a) shows the $h$ dependence of $D_{AB}$ and $C_{AB}$, which are crucial factors for positive $\Delta E_{B}^{\rm max}$ and $\Delta E_{B,B}^{\rm max}$, respectively, as already shown in Eqs.~(\ref{DEBmax}) and (\ref{DEBBmax}). These correlators are large for a small-$h$ region. Their behaviors after multiplying $h$ show only weak-$h$ dependence except for the small-$h$ region. Figure~\ref{KitaevQETfig3}(b) shows $\Delta E_{A}$ for $\vec{r}=(0,1,0)$ and $\vec{r}=(1,0,0)$, corresponding to $\Delta E_{B}^{\rm max}$ and $\Delta E_{B,B}^{\rm max}$, respectively, as a function of $h$. We first examine the $h$ dependence of $\Delta E_{A}$. The local energy change by the projection measurement is mediated by the Alice's local Hamiltonian $H_{A}$. It is natural that $\Delta E_{A}$ increases as $h$ increases. We next examine the $h$ dependence of $\Delta E_{B}^{\rm max}$ and $\Delta E_{B,B}^{\rm max}$. Both of them are zero for $h=0$, and start to increase with $h$. Further increase of $h$ reduces their magnitudes, because the effect of $hD_{AB}$ and $hC_{AB}$ on $\Delta E_{B}^{\rm max}$ and $\Delta E_{B,B}^{\rm max}$ is smeared out by the increase of $\left|\epsilon_{B}\right|$ in large-$h$ region. We find $\Delta E_{A}>\Delta E_{B}^{\rm max}$ for $r_{y}=1$. The ratio, $\Delta E_{B}^{\rm max}/\Delta E_{A}$ is about $0.032$ for $h=0.18k$ that is a peak position of $\Delta E_{B}^{\rm max}$. We also find $\Delta E_{A}>\Delta E_{B,B}^{\rm max}$ for $r_{x}=1$ and $h>0.24k$. The inequality is reversed for $h<0.24k$. We plot $\Delta E_{B}$ as the green dash-dotted line in Fig.~\ref{KitaevQETfig3}(b) when $\Delta E_{B,B}$ is maximized. This quantity is given by the sum of Eqs.~(\ref{DEBBmax}) and (\ref{debin}). We find that the quantity is negative and $\left|\Delta E_{B,R}\right|\gg\Delta E_{B,B}^{\rm \max}$. Therefore, even if the amount of reduction in the internal energy for $h<0.24k$ is greater than the amount of energy injected into the system, the work to be taken out of the system does not exceed the amount of energy injected.

\section{Majorana Fermion Representation of Nonlocal Correlators}
\label{Majorana}

Our main purpose of this paper is to convert the present QET model into a Majorana fermion model so that we examine relationship between QET and Majorana nonlocal correlation. The conversion is realized by introducing the following Majorana representation:
\begin{align}
b_{l}&=f_{l}^{\dagger}+f_{l} , \\
c_{l}&=i\bigl(f_{l}^{\dagger}-f_{l}\bigr) , \\
c_{m}&=f_{m}^{\dagger}+f_{m} , \\
b_{m}&=i\bigl(f_{m}^{\dagger}-f_{m}\bigr) ,
\end{align}
for $l=A,C_{2}$ and $m=C_{1},B$. Then, the QET Hamiltonian is mapped onto the following form:
\begin{eqnarray}
H=ihb_{A}c_{A}-ik\left(c_{A}c_{C_{1}}-c_{C_{1}}c_{C_{2}}+c_{C_{2}}c_{B}\right)+ihc_{B}b_{B}. \label{KH} \nonumber \\
\end{eqnarray}
For $h=0$, the Hamiltonian is described by the bilinear form of itinerant $c$-Majorana fermions, and $b$-Majorana fermions are zero modes. The introduction of a finite $h$ value induces coupling between $c$- and $b$-Majorana fermions at both edges. Let us compare this model with the conventional Kitaev model defined on a two-dimensional honeycomb lattice: $H=\sum_{\left<i,j\right>_{a}}J_{a}\sigma_{i}^{a}\sigma_{j}^{a}$, where $\left<i,j\right>_{a}$ represents the sum over nearest-neighbor pairs connected by an $a$-bond (each vertex in the honeycomb lattice has three bonds, and we label them $a=x,y,z$). By the transformation $\sigma_{j}^{a}=ib_{j}^{a}c_{j}$, the Hamiltonian is mapped onto an itinerant $c$-Majorana fermion system coupled to $Z_{2}$ gauge fields $u_{ij}^{a}=ib_{i}^{a}b_{j}^{a}$. The introduction of magnetic field induces coupling between $b$- and $c$-Majorana fermions. Although the transformation from spin operators to Majorana operators in our model is different from that in the Kitaev model, a correspondence between these two models can be observed as follows: the Majorana hopping with $J_{a}$ in the Kitaev model corresponds to the $k$ term in Eq.~(\ref{KH}). The Zeeman-field term in the Kitaev model is usually represented as $-\sum_{a}h_{a}\sigma_{j}^{a}=-\sum_{a}ih_{a}b_{j}^{a}c_{j}$ at site $j$, and this representation corresponds to the edge-field term in Eq.~(\ref{KH}). Thus, our model can be viewed as one-dimensional simplification of the Kitaev model with edge magnetic fields~\cite{Minakawa,Koga,Taguchi,Nasu,Misawa}. In the previous work~\cite{Takahashi}, the Kitaev model with two vacancies under a magnetic field is examined. In this case also, the $b$-$c$ hopping terms of the nearest-neighbor sites of the vacancies appear, when we treat the magnetic field perturbatively.

We focus on the Majorana correlation, $\left<\psi\right|ib_{A}b_{B}\left|\psi\right>$, that is evaluated as
\begin{align}
\left<\psi\right|ib_{A}b_{B}\left|\psi\right>=-4Z^{2}\left(1+\alpha\beta\right), \label{bAbBeven}
\end{align}
which reminds us of $C_{AB}$ in Eq.~(\ref{CAB}). In order to clarify the reason for their equivalence except for the sign, the operator product $ib_{A}b_{B}$ before taking the expectation value is evaluated as
\begin{align}
ib_{A}b_{B}&=\sigma_{A}^{x}\left(\sigma_{A}^{z}\sigma_{C_{1}}^{z}\sigma_{C_{2}}^{z}\right)\left(i\sigma_{B}^{y}\right)=\sigma_{A}^{x}\sigma_{B}^{x}(-P), \label{bAbB}
\end{align}
where $P$ ($=\sigma_{A}^{z}\sigma_{C_{1}}^{z}\sigma_{C_{2}}^{z}\sigma_{B}^{z}$) is the parity operator. For the even-parity state $\left|\psi\right>$ ($P\left|\psi\right>=\left|\psi\right>$), we actually obtain
\begin{align}
\left<\psi\right|ib_{A}b_{B}\left|\psi\right>=\left<\psi\right|\sigma_{A}^{x}\sigma_{B}^{x}(-1)\left|\psi\right>=-C_{AB}. \label{BC}
\end{align}
For the odd-parity sector, see Appendix~\ref{appC}. We also consider the Majorana representation of $D_{AB}$ and $C_{AR}$. The results are given by
\begin{align}
D_{AB}&=\left<\psi\right|ic_{A}c_{B}P\left|\psi\right>=\left<\psi\right|ic_{A}c_{B}\left|\psi\right> , \\
C_{AR}&=\left<\psi\right|ib_{A}c_{C_{2}}P\left|\psi\right>=\left<\psi\right|ib_{A}c_{C_{2}}\left|\psi\right>.
\end{align}
The quantity $C_{AB}$ ($D_{AB}$) represents correlation between the $b$-Majorana ($c$-Majorana) fermions at both edges. These correlators, $C_{AB}$ and $D_{AB}$, appear as a result of feedback control after classical communication between Alice and Bob, and are important for maximizing $\Delta E_{B,B}$ and $\Delta E_{B}$, respectively. On the other hand, the correlator $C_{AR}$ in $\Delta E_{B,R}$ is less important for the energy extraction. In particular, $\Delta E_{B,R}$ disappears for the maximization of $\Delta E_{B}$. The Majorana representation of $C_{AR}$ does not contain information about Majorana fermions at Bob's site, even though the original representation in Eq.~(\ref{DAB}) contains the spin operator $\sigma_{B}^{z}$ at Bob's site. Therefore, we consider that the Majorana representation is a nice benchmark of how the communication between Alice and Bob is 
reflected to the maximum energy extraction and the maximum local-energy reduction.

\section{On the First and Second Laws of Effective Information Thermodynamics}
\label{IT}

Let us discuss effective information-thermodynamical aspects behind the present QET model. Equation~(\ref{BBI}) is suggestive for considering the heat transfer that inevitably occurs in the QET process. The quantity $\Delta E_{B}$ is work that is taken out of the whole four-site system by Bob's local feedback control. Let us denote the work that Bob's local system gets as $W=-\Delta E_{B}$ and the increase in the internal energy of Bob's local system as $\Delta U=-\Delta E_{B,B}$. Then, Eq.~(\ref{BBI}) is reduced to the first law of effective thermodynamics for Bob's local system, $W+Q=\Delta U$, where $Q$ is the heat absorbed by Bob's local system and assigned as
\begin{align}
Q=\Delta E_{B,R}.
\end{align}
For the negative $\Delta E_{B,R}$, heat is necessarily released to the outside of Bob's local system through the interaction term $H_{R}$. A special case is maximization of $\Delta E_{B}$, in which the heat transfer does not occur.

Equation~(\ref{DEBB}) (or (\ref{DEBBmax})) reminds us of the upper bound of work extraction in the second law of information thermodynamics, since the nonlocal correlator $C_{AB}$, quantum information in the ground state, is a crucial source for maximizing $\Delta E_{B,B}$. However, the nonlocal correlator is different from the QC-mutual information that is a resource of the work extraction. In this section, we discuss their difference.

The second law is usually defined for isothermal processes at temperature $T$, and is given by $W_{\rm ext}\le -\Delta F+T I_{\rm QC}$, where the upper bound of the work extraction $W_{\rm ext}$ contains information-gain term characterized by the QC-mutual information $I_{\rm QC}$ as well as free-energy change $\Delta F$~\cite{Sagawa,Groenewold,Ozawa}. Motivated by the second law, some of the present authors have recently derived the upper bound on locally extractable energy from entangled pure states under feedback control~\cite{Itoh2}. The present setup matches with that of Ref.~\cite{Itoh2}, when we regard $H_{B}$ as the Hamiltonian of the local system of our interest (in Ref.~\cite{Itoh2}, we did not consider interaction terms and there is no heat transfer, even if we consider the maximization of $\Delta E_{B,B}$). The energy reduction $\Delta E_{B,B}$ shows the second-law-like behavior as
\begin{align}
\Delta E_{B,B}\le {\cal F}(\rho_{B}^{\rm i};H_{B})-F(\sigma_{B})+\frac{1}{\beta_{\rm eff}}I_{\rm QC}, \label{Kanji}
\end{align}
where ${\cal F}(\rho_{B}^{\rm i};H_{B})$ is the nonequilibrium free energy for the reduced density matrix of the ground state $\rho_{B}^{\rm i}={\rm tr}_{AC_{1}C_{2}}\left|\psi\right>\left<\psi\right|$, $F(\sigma_{B})$ is the free energy of the thermal equilibrium state for the local Hamiltonian $H_{B}$, $\sigma_{B}=e^{-\beta_{\rm eff}H_{B}}/Z_{\rm eff}$ with the partition function $Z_{\rm eff}$, and $\beta_{\rm eff}$ is an effective inverse temperature that originates in the coupling between $B$ site and $C_{1}C_{2}$ sites. Two important quantites in Eq.~(\ref{Kanji}) are the free-energy difference and the QC-mutual information, $I_{\rm QC}$, respectively defined by
\begin{align}
&{\cal F}(\rho_{B}^{\rm i};H_{B})-F(\sigma_{B})=\frac{1}{\beta_{\rm eff}}D\left(\rho_{B}^{\rm i}||\sigma_{B}\right)\ge 0, \label{KL} \\
&I_{\rm QC}=S\left(\rho_{B}^{\rm i}\right)-\sum_{n=\pm 1}p_{n}S\left(\rho_{B}^{\rm m}(n)\right)\ge 0 , \label{IQC}
\end{align}
where $D\left(\rho_{B}^{\rm i}||\sigma_{B}\right)={\rm tr}_{B}\left(\rho_{B}^{\rm i}\log\left(\rho_{B}^{\rm i}/\sigma_{B}\right)\right)$ denoting the Kullback-Leibler (KL) divergence, $S(\sigma)=-{\rm tr}_{B}\left(\sigma\log\sigma\right)$ denoting the von Neumann entropy of a density matrix $\sigma$, and $\rho_{B}^{\rm m}(n)={\rm tr}_{AC_{1}C_{2}}\left|\psi_{A}(n)\right>\left<\psi_{A}(n)\right|$. The KL divergence shows how the initial nonequilibrium state is far from $\sigma_{B}$, and is also an important resource of the energy reduction as well as the QC-mutual information. The equality condition of Eq.~(\ref{Kanji}) holds only if the following relation is satisfied:
\begin{align}
S(\sigma_{B})=\min_{\left\{P_{A}(n)\right\}}\sum_{n}p_{n}S\left(\rho_{B}^{\rm m}(n)\right), \label{minimization}
\end{align}
where the minimization is performed over all possible projection operators, and is realized for $\vec{r}=(1,0,0)$ (see Appendix~\ref{appD}). The projection operator $P_{A}(n)$ for the minimization also gives the maximum value of $\Delta E_{B,B}$ in the present case. Equation (\ref{minimization}) is used for the determination of $\beta_{\rm eff}$.

As shown in Appendix~\ref{appE}, the right-hand side (RHS) of Eq.~(\ref{Kanji}) becomes equivalent to $\Delta E_{B,B}^{\rm max}$ in Eq.~(\ref{DEBBmax}) for $\beta_{\rm eff}$ determined by Eq.~(\ref{minimization}). According to Eq.~(\ref{DEBB}), $\Delta E_{B,B}^{\rm max}$ is the sum of $\epsilon_{B}\left(1-\cos(2\theta)\right)$ and $-hC_{AB}\sin(2\theta)$ with Eqs.~(\ref{sine}) and (\ref{cosine}). The former contribution, $\epsilon_{B}\left(1-\cos(2\theta)\right)$, is always negative, and then the latter contribution, $-hC_{AB}\sin(2\theta)$ is larger than the RHS of Eq.~(\ref{Kanji}). The KL divergence contains information about entanglement-entropy change by the measurement, and thus not only the QC-mutual information but also the KL divergence is related to $C_{AB}$.

\begin{figure}[htbp]
\begin{center}
\includegraphics[width=7cm]{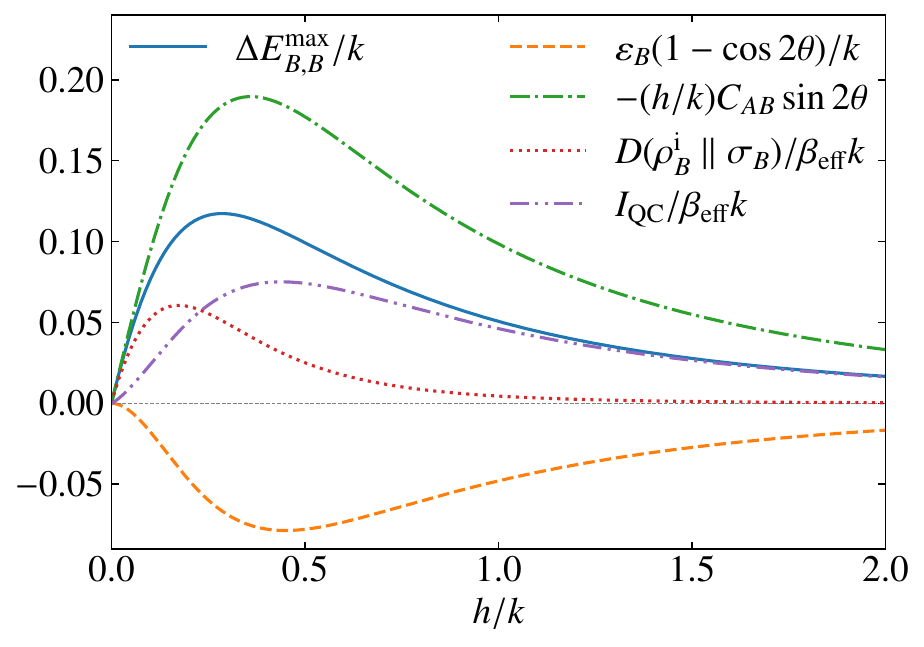}
\end{center}
\caption{$\Delta E_{B,B}^{\rm max}$, $\epsilon_{B}\left(1-\cos(2\theta)\right)$, $-hC_{AB}\sin(2\theta)$, $D\left(\rho_{B}^{\rm i}||\sigma_{B}\right)/\beta_{\rm eff}$, and $I_{\rm QC}/\beta_{\rm eff}$ as a function of $h$. The trigonometric functions are defined by Eqs.~(\ref{sine}) and (\ref{cosine}).
}
\label{KitaevQETfig4}
\end{figure}

Figure~\ref{KitaevQETfig4} shows the $h$ dependence of all components in $\Delta E_{B,B}^{\rm max}$ and the RHS of Eq.~(\ref{Kanji}). We actually observe $-hC_{AB}\sin(2\theta)\ge\Delta E_{B,B}^{\rm max}=\left(D\left(\rho_{B}^{\rm i}||\sigma_{B}\right)+I_{\rm QC}\right)/\beta_{\rm eff}$. As $h$ increases, $D\left(\rho_{B}^{\rm i}||\sigma_{B}\right)/\beta_{\rm eff}$ tends to disappear. The decay of $D\left(\rho_{B}^{\rm i}||\sigma_{B}\right)/\beta_{\rm eff}$ is more remarkable than that of $I_{\rm QC}/\beta_{\rm eff}$. For $h/k>1$, the RHS of Eq.~(\ref{Kanji}) is almost equal to $I_{\rm QC}/\beta_{\rm eff}$. In this parameter region, $\epsilon_{B}\left(1-\cos(2\theta)\right)$ still remains, and this decay is slow. These data show that the nonlocal correlation term is not simply equal to the QC-mutual information even for the parameter region in which the KL divergence disappears.

\section{Concluding Remarks}
\label{conclusion}

We have considered QET on a quantum spin model and its Majorana representation in order to examine the roles of the Majorana correlations on the teleportation properties. We have derived the formula for connecting the maximally extracted energy $\Delta E_{B}^{\rm max}$ with the nonlocal $c$-Majorana correlator $D_{AB}$, and have shown that the correlator is necessary for the positive energy extraction. We have also derived the formula about the maximum of the local energy reduction $\Delta E_{B,B}^{\rm max}$ that is consistent with our previous work~\cite{Itoh2}. In this case, the nonlocal $b$-Majorana correlator $C_{AB}$ is necessary for the energy reduction. In both cases, the nonlocal correlators appear as a result of Bob's feedback control. We have assigned $\Delta E_{B,R}$ as an effective heat absorption by the local system. We have found that there is no heat transfer when the extracted energy is maximized. We have also found that the Majorana correlator $C_{AB}$ is different from the QC-mutual information in our previous work~\cite{Itoh2}, although both of the Majorana correlator and the QC-mutual information are quantum-information resources for driving these protocols.

Equations (\ref{BBI}) and (\ref{DEBi}) hold regardless of the details of $H_{C}$, although the values of $C_{AB}$, $D_{AB}$, and $C_{AR}$ depend on the system size $L$. Thus, our statements are applicable to larger systems and systems with different geometries and spatial dimension. We have numerically evaluated those correlators for $H_{C}=-ik\sum_{l=0}^{L-2}(-1)^{l}c_{l}c_{l+1}$ up to $L=1000$ in one dimension. For various $L$ values, we have found that $C_{AB}=1$ for $h=0$ and $C_{AB}$ decreases with increasing $h$ (see also Fig.~\ref{KitaevQETfig3}(a) for $L=4$). With increasing $L$, the decrease becomes stronger. For various $L$ values, we have also found that the absolute value of $D_{AB}$ is maximum for $h=0$ and decreases with increasing $h$ (see also Fig.~\ref{KitaevQETfig3}(a) for $L=4$). The absolute value shows power-law decay as a function of $L$ for finite-$h$ values, which is an evidence of quantum criticality of our model. The slow decay due to the power-law correlation enables us to consider QET for relatively large-$L$ systems.

In the previous study of the Kitaev model~\cite{Minakawa}, the induced magnetic moment after the pulse field excitation is measured by the time-dependent calculation. The time delay of the classical communication in the QET protocol may weaken the QET performance. It is a future work whether our statement holds in time-dependent cases.

This work was supported by JSPS KAKENHI Grant Numbers JP24K06878, JP24K00563, JP24K02948, JP23K22492, and CSIS in Tohoku University.

\appendix
\section{Symmetry operations}
\label{appA}

The eigenstates of our model can be characterized by various operators commuting or anticommuting with the Hamiltonian. We define the following four operators:
\begin{align}
P&=\sigma_{A}^{z}\sigma_{C_{1}}^{z}\sigma_{C_{2}}^{z}\sigma_{B}^{z} , \\
Q&=\sigma_{A}^{z}\sigma_{C_{1}}^{z}\sigma_{C_{2}}^{x} , \\
R&=\sigma_{C_{1}}^{x}\sigma_{C_{2}}^{x} , \\
S&=\sigma_{A}^{x}\sigma_{C_{1}}^{y}\sigma_{C_{2}}^{x}\sigma_{B}^{y} ,
\end{align}
where we find $\left[P,H\right]=0$, $\left[Q,H\right]=0$, $\left\{P,Q\right\}=0$, $\left[R,H\right]=0$, $\left[R,P\right]=0$, $\left\{R,Q\right\}=0$, $\left\{S,H\right\}=0$, $\left[P,S\right]=0$, $P^{2}=1$, $Q^{2}=1$, $R^{2}=1$, and $S^{2}=1$. The operator $P$ is the parity operator for the number of Jordan-Wigner fermions with eigenvalues $\pm 1$, and the total Hilbert space is devided into two subspaces with even or odd numbers of Jordan-Wigner fermions. The different-parity states degenerate. We only treat the eigenstates $\left|\Psi_{+}\right>$ in the even-parity sector ($H\left|\Psi_{+}\right>=E\left|\Psi_{+}\right>$, $P\left|\Psi_{+}\right>=\left|\Psi_{+}\right>$) in the main text. The operator $R$ has the eigenvalues $\pm 1$ (we shortly call it as $R$-parity), and can be efficiently used for identifying the eigenstates. The operator $S$ corresponds to spin inversion, and the spectra are symmetric with respect to the negative and positive sides in a specific parity sector.

The Jordan-Wigner representation of the operator $R$ is given by $R=\sigma_{C_{1}}^{x}\sigma_{C_{2}}^{x}=\bigl(f_{C_{1}}^{\dagger}-f_{C_{1}}\bigr)\bigl(f_{C_{2}}^{\dagger}+f_{C_{2}}\bigr)$. The eigenstates of $R$ are summarized as follows:
\begin{align}
R\left(\left|eeff\right>\pm\left|efef\right>\right)&=\pm\left(\left|eeff\right>\pm\left|efef\right>\right) , \\
R\left(\left|ffee\right>\pm\left|fefe\right>\right)&=\pm\left(\left|ffee\right>\pm\left|fefe\right>\right) , \\
R\left(\left|ffff\right>\pm\left|feef\right>\right)&=\pm\left(\left|ffff\right>\pm\left|feef\right>\right) , \\
R\left(\left|eeee\right>\pm\left|effe\right>\right)&=\pm\left(\left|eeee\right>\pm\left|effe\right>\right) ,
\end{align}
where we take the duplicate signs in the same order. The general eigenstates in the even-parity sector can be represented as
\begin{align}
\left|\Psi_{+}\right>=&a_{1}\left(\left|eeff\right>+r\left|efef\right>\right)+a_{2}\left(\left|ffff\right>+r\left|feef\right>\right) \nonumber \\
&+\lambda a_{1}\left(\left|ffee\right>+r\left|fefe\right>\right)+a_{3}\left(\left|eeee\right>+r\left|effe\right>\right) ,
\end{align}
where $r$ ($=\pm 1$) is the eigenvalue of $R$, $\lambda$ ($=\pm 1$) is the eigenvalue of space inversion, and parameters $a_{i}$ ($i=1,2,3$) are determined by solving $H\left|\Psi_{+}\right>=E\left|\Psi_{+}\right>$ and by the normalization condition (we do not explicitly introduce the space-inversion operator, since it is not easy to write it by the Pauli matrices).

Furthermore, we introduce the operator
\begin{align}
Q\sigma_{A}^{x}=i\sigma_{A}^{y}\sigma_{C_{1}}^{z}\sigma_{C_{2}}^{x} . \label{QbA}
\end{align}
This operator commutes with $H$ for $h=0$. A remarkable property of this operator is to satisfy $\left\{Q\sigma_{A}^{x},R\right\}=0$ and $\left[Q\sigma_{A}^{x},P\right]=0$. The state $Q\sigma_{A}^{x}\left|\Psi_{+}\right>$ degenerates with $\left|\Psi_{+}\right>$, but the eigenvalue of $R$ for $Q\sigma_{A}^{x}\left|\Psi_{+}\right>$ is different from that for $\left|\Psi_{+}\right>$. Thus, $R$ identifies the degeneracy at $h=0$. Because of the $R$-parity change, we obtain
\begin{align}
\left<\Psi_{+}\right|Q\sigma_{A}^{x}\left|\Psi_{+}\right>=0. \label{Rparity}
\end{align}

\section{Derivation of $\Delta E_{B}^{\rm max}$ and $\Delta E_{B,B}^{\rm max}$}
\label{appB}

Let us derive the maximum of $\Delta E_{B}$, $\Delta E_{B}^{\rm max}$. For this purpose, we first transform $\Delta E_{B}$ into the following form:
\begin{align}
\Delta E_{B}=\sqrt{W_{0}^{2}+X_{0}^{2}}\cos\left(2\theta+\delta_{0}\right)-\left|W_{0}\right| ,
\end{align}
where
\begin{align}
W_{0}&=\left(1-s_{z}^{2}\right)\epsilon_{B}+\left(1-s_{x}^{2}\right)\epsilon_{R}, \\
X_{0}&=r_{x}s_{y}\left(kC_{AR}-hC_{AB}\right)+r_{y}s_{x}hD_{AB},
\end{align}
and the phase factor $\delta_{0}$ is determined by the following relations:
\begin{align}
\cos\delta_{0}&=\frac{-W_{0}}{\sqrt{W_{0}^{2}+X_{0}^{2}}} , \\
\sin\delta_{0}&=\frac{-X_{0}}{\sqrt{W_{0}^{2}+X_{0}^{2}}}.
\end{align}
We find
\begin{align}
\Delta E_{B}\le\sqrt{W_{0}^{2}+X_{0}^{2}}-\left|W_{0}\right|, \label{DEB}
\end{align}
and $\Delta E_{B}^{\rm max}$ is obtained for $2\theta+\delta_{0}=0$. The RHS of Eq.~(\ref{DEB}) is a function of $\vec{r}$ and $\vec{s}$, which are determined to maximize $\Delta E_{B}$. The factor $W_{0}$ is represented as
\begin{align}
W_{0}=\epsilon_{B}\sin^{2}\xi+\epsilon_{R}\left(1-\sin^{2}\xi\cos^{2}\eta\right).
\end{align}
According to Eq.~(\ref{ab}), we find
\begin{align}
kC_{AR}-hC_{AB}=-hD_{AB}, \label{CCD}
\end{align}
and thus the factor $X_{0}$ is represented as
\begin{align}
X_{0}&=\left(r_{y}s_{x}-r_{x}s_{y}\right)h D_{AB} \nonumber \\
&=hD_{AB}\sin\mu\sin\xi\sin\left(\nu-\eta\right).
\end{align}
The RHS of Eq.~(\ref{DEB}) is an increasing function of $\sin^{2}\xi$, and is maximized for $\xi=\pi/2$. 
After that, we obtain $W_{0}=\epsilon_{B}+\epsilon_{R}\sin^{2}\eta$ and $X_{0}=hD_{AB}\sin\mu\sin\left(\nu-\eta\right)$. The $\eta$ dependence of $X_{0}$ can be absorbed by the optimization of $\nu$, and thus, we can determine $\eta$ to optimize $W_{0}$. The RHS of Eq.~(\ref{DEB}) is a decreasing function of $\left|W_{0}\right|$. Since $\epsilon_{B}$ and $\epsilon_{R}$ are negative, we take $\eta=0$. The RHS of Eq.~(\ref{DEB}) is an increasing function of $\left|X_{0}\right|$, and thus we take $\mu=\nu=\pi/2$. Finally, we obtain Eq.~(\ref{DEBmax}).

We also derive the maximum value of $\Delta E_{B,B}$, $\Delta E_{B,B}^{\rm max}$. For this purpose, we transform $\Delta E_{B,B}$ into the following form:
\begin{align}
\Delta E_{B,B}=\sqrt{W^{2}+X^{2}}\cos\left(2\theta+\delta\right)-\left|W\right|,
\end{align}
where
\begin{align}
W&=\epsilon_{B}\left(1-s_{z}^{2}\right)=\epsilon_{B}\sin^{2}\xi , \\
X&=-h\left(r_{x}s_{y}C_{AB}-r_{y}s_{x}D_{AB}\right) \nonumber \\
&=-4Z^{2}h\sin\mu\sin\xi\left\{\sin\left(\eta-\nu\right)+\alpha\beta\sin\left(\eta+\nu\right)\right\},
\end{align}
and the phase factor $\delta$ is determined from the the following relations:
\begin{align}
\cos\delta &=\frac{-W}{\sqrt{W^{2}+X^{2}}} , \\
\sin\delta &=\frac{-X}{\sqrt{W^{2}+X^{2}}}.
\end{align}
We find
\begin{align}
\Delta E_{B,B}\le\sqrt{W^{2}+X^{2}}-\left|W\right|,
\label{DEBB2}
\end{align}
and $\Delta E_{B,B}^{\rm max}$ is obtained for $2\theta+\delta=0$. The RHS of Eq.~(\ref{DEBB2}) is to be maximized with respect to $\vec{r}$ and $\vec{s}$. The RHS of Eq.~(\ref{DEBB2}) is an increasing function of $\sin^{2}\xi$, and we take $\xi=\pi/2$. Then, the RHS of Eq.~(\ref{DEBB2}) is maximized for the maximum value of $\left|X\right|$. We take $\mu=\eta=\pi/2$ and $\nu=0$, since $\alpha\beta>0$. Finally, we obtain Eq.~(\ref{DEBBmax}).

\section{Effect of degeneracy between even- and odd-parity states}
\label{appC}

In the main text, we focused on the lowest-energy state $\left|\psi\right>$ in the even-parity sector. The state $\left|\psi\right>$ degenerates with $\left|\phi\right>=Q\left|\psi\right>$ in the odd-parity sector. Here, we check whether our result holds in the odd-parity sector. For this purpose, we can use the following transformation [$P_{A}(n)=\left(I_{A}+n\vec{r}\cdot\vec{\sigma}_{A}\right)/2$]:
\begin{align}
&QP_{A}(n)Q=P_{A}(-n) \;\;\;\;\left(r_{z}=0\right) , \label{plusminus} \\
&QU_{B}(n)Q=U_{B}(n) , \\
&Q\left(ib_{A}b_{B}\right)Q=ib_{A}b_{B} , \\
&Q\left(ic_{A}c_{B}\right)Q=ic_{A}c_{B} ,
\end{align}
where $Qb_{i}Q=-b_{i}$ and $Qc_{i}Q=-c_{i}$ ($i=A,B$). We find
\begin{align}
&\left<\phi\right|ib_{A}b_{B}\left|\phi\right>=\left<\psi\right|ib_{A}b_{B}\left|\psi\right>, \label{evenodd} \\
&\left<\phi\right|ic_{A}c_{B}\left|\phi\right>=\left<\psi\right|ic_{A}c_{B}\left|\psi\right>.
\end{align}
Thus, the $b$- and $c$-Majorana correlators are equivalent in both even- and odd-parity sectors. We also find that $Q$ changes the sign of $n$ in Eq.~(\ref{plusminus}). The QET performance at least for $r_{z}=0$ is not affected by the parity change, since the sign change in Eq.~(\ref{plusminus}) can be absorbed by that of $r_{x}$ and $r_{y}$.

It would be helpful for readers to mention four-fold degeneracy of the lowest-energy eigenstates for $h=0$. The eigenstates are identified with the signs of the eigenvalues of $P$ and $R$ where the eigenvalue of $P$ is denoted as $p$ ($=\pm 1$). The eigenstates are specified as $\left|+,+\right>=\left|\psi\right>$, $\left|+,-\right>=Q\sigma_{A}^{x}\left|\psi\right>$, $\left|-,+\right>=b_{A}\left|\psi\right>$, and $\left|-,-\right>=Q\left|\psi\right>=\left|\phi\right>$, where the two signs in $\left|\pm,\pm\right>$ correspond to the signs of $p$ and $r$ from the left. Then, we find
\begin{align}
&\left<p,r\right|ib_{A}b_{B}\left|p,r\right>=-prC_{AB}, \\
&\left<p,r\right|ic_{A}c_{B}\left|p,r\right>=D_{AB}.
\end{align}
Therefore, at least for $h=0$, the absolute values of the $b$- and $c$-Majorana correlators do not change for any combination of $p$ and $r$.

\section{On the minimization in Eq.~(\ref{minimization})}
\label{appD}

We show some details about the minimization in Eq.~(\ref{minimization}). For general $P_{A}(n)=(I_{A}+n\vec{r}\cdot\vec{\sigma}_{A})/2$, the reduced density matrix $\rho_{B}^{\rm m}(n)$ is given by
\begin{align}
\rho_{B}^{\rm m}(n)=&\left(1-\theta\right)\left|e\right>\left<e\right|+\kappa\left|e\right>\left<f\right|+\kappa^{\ast}\left|f\right>\left<e\right| \nonumber \\
&+\theta\left|f\right>\left<f\right| ,
\end{align}
where $\theta$ and $\kappa$ are respectively defined by
\begin{align}
\theta&=\frac{2Z^{2}\left(\left(1-nr_{z}\right)+\alpha^{2}\left(1+nr_{z}\right)\right)}{1+2nr_{z}\left(\alpha^{2}-\beta^{2}\right)Z^{2}}, \\
\kappa&=\frac{2nZ^{2}\left(\left(r_{x}+ir_{y}\right)+\alpha\beta\left(r_{x}-ir_{y}\right)\right)}{1+2nr_{z}\left(\alpha^{2}-\beta^{2}\right)Z^{2}} .
\end{align}
The eigenvalues of $\rho_{B}^{\rm m}(n)$ are given by
\begin{align}
\lambda_{\pm}^{\rm m}(n)&=\frac{1}{2}\left(1\pm\sqrt{1-4\theta\left(1-\theta\right)+4\left|\kappa\right|^{2}}\right).
\end{align}
We find that $\lambda_{\pm}^{\rm m}(n)$ depend on both of $n$ and $\vec{r}$. Furthermore, we find
\begin{align}
p_{n}=\frac{1}{2}\left(1+2nr_{z}\left(\alpha^{2}-\beta^{2}\right)Z^{2}\right),
\end{align}
and $p_{n}$ also depends on $n$. We examine $\sum_{n}p_{n}S\left(\rho_{B}^{\rm m}(n)\right)$ as functions of $h/k$ and $\vec{r}$. We have confirmed numerically that $\sum_{n}p_{n}S\left(\rho_{B}^{\rm m}(n)\right)$ is minimized for $r_{x}=1$. Therefore, it is reasonable to take $P_{A}(n)=\left(I_{A}+n\sigma_{A}^{x}\right)/2$.

\section{Proof of equivalence between Eq.~(\ref{DEBBmax}) and the RHS of Eq.~(\ref{Kanji}) for optimized $P_{A}(n)$}
\label{appE}

The equivalence between Eq.~(\ref{DEBBmax}) and the RHS of Eq.~(\ref{Kanji}) for optimized $P_{A}(n)$ is derived by using $\beta_{\rm eff}$ that is a unique solution of Eq.~(\ref{minimization}).

For the derivation of $\beta_{\rm eff}$, let us start with $\rho_{B}^{\rm i}$ and $\rho_{B}^{\rm m}(n)$ that are respectively evaluated as
\begin{align}
\rho_{B}^{\rm i}=&2Z^{2}\left[ \left(1+\beta^{2}\right)\left|e\right>\left<e\right| +\left(1+\alpha^{2}\right)\left|f\right>\left<f\right| \right] , \\
\rho_{B}^{\rm m}(n)=&2Z^{2}\left[ \left(1+\beta^{2}\right)\left|e\right>\left<e\right| +\left(1+\alpha^{2}\right)\left|f\right>\left<f\right| \right. \nonumber \\
& \left. +n\left(1+\alpha\beta\right)\left(\left|e\right>\left<f\right|+\left|f\right>\left<e\right|\right)\right] .
\end{align}
A remarkable property of $\rho_{B}^{\rm m}(n)$ is that the eigenvalues, $\lambda_{\pm}^{\rm m}$, do not depend on $n$:
\begin{align}
\lambda_{\pm}^{\rm m}=\frac{1}{2}\left(1\pm\sqrt{1-16Z^{4}\left(\alpha-\beta\right)^{2}}\right)\equiv\frac{1}{2}\left(1\pm g\right). \label{lambdam}
\end{align}
Thus, $I_{\rm QC}$ is simply represented as $I_{\rm QC}=S(\rho_{B}^{\rm i})-S(\rho_{B}^{\rm m}(n))$. Since $\lambda_{\pm}^{\rm m}$ is independent of $n$, Eq.~(\ref{minimization}) is reduced to $S(\sigma_{B})=S\left(\rho_{B}^{\rm m}(n)\right)$ for the optimized $P_{A}(n)$. Then, we can solve Eq.~(\ref{minimization}) for $\beta_{\rm eff}$ by just comparing the eigenvalues of $\sigma_{B}$ with $\lambda_{\pm}^{\rm m}$. We find
\begin{align}
\beta_{\rm eff}h=\log\sqrt{\lambda_{+}^{\rm m}/\lambda_{-}^{\rm m}}. \label{beta}
\end{align}

Let us compare $\Delta E_{B,B}^{\rm max}$ in Eq.~(\ref{DEBBmax}) with the RHS of Eq.~(\ref{Kanji}). For this purpose, we transform the RHS into the following form:
\begin{align}
&{\cal F}\left(\rho_{B}^{\rm i};H_{B}\right)-F\left(\sigma_{B}\right)+\frac{1}{\beta_{\rm eff}}I_{\rm QC} \nonumber \\
&=\epsilon_{B}+\frac{h}{\log\sqrt{\lambda_{+}^{\rm m}/\lambda_{-}^{\rm m}}}\left(\log Z_{\rm eff}-S\left(\rho_{B}^{\rm m}\right)\right),
\end{align}
where the partition function is evaluated as $Z_{\rm eff}={\rm tr}_{B}e^{-\beta_{\rm eff}H_{B}}=1/\sqrt{\lambda_{+}^{\rm m}\lambda_{-}^{\rm m}}$. We can analytically show
\begin{align}
&\sqrt{\left(\frac{\epsilon_{B}}{h}\right)^{2}+C_{AB}^{2}} \nonumber \\
&=\sqrt{\left(2Z^{2}\left(\alpha^{2}-\beta^{2}\right)^{2}\right)^{2}+\left(4Z^{2}\left(1+\alpha\beta\right)\right)^{2}} \nonumber \\
&=2Z^{2}\sqrt{4+8\alpha\beta+\left(\alpha^{2}+\beta^{2}\right)^{2}} \nonumber \\
&=g,
\end{align}
and
\begin{align}
&\frac{\log Z_{\rm eff}-S\left(\rho_{B}^{\rm m}(n)\right)}{\log\sqrt{\lambda_{+}^{\rm m}/\lambda_{-}^{\rm m}}} \nonumber \\
&=\frac{-\log\sqrt{\lambda_{+}^{\rm m}\lambda_{-}^{\rm m}}+\lambda_{+}^{\rm m}\log\lambda_{+}^{\rm m}+\lambda_{-}^{\rm m}\log\lambda_{-}^{\rm m}}{\log\sqrt{\lambda_{+}^{\rm m}/\lambda_{-}^{\rm m}}} \nonumber \\
&=\frac{-\log\sqrt{\lambda_{+}^{\rm m}\lambda_{-}^{\rm m}}+\frac{1}{2}(1+g)\log\lambda_{+}^{\rm m}+\frac{1}{2}(1-g)\log\lambda_{-}^{\rm m}}{\log\sqrt{\lambda_{+}^{\rm m}/\lambda_{-}^{\rm m}}} \nonumber \\
&=g.
\end{align}
Therefore, the equality between $\Delta E_{B,B}^{\rm max}$ in Eq.~(\ref{DEBBmax}) and the RHS of Eq.~(\ref{Kanji}) is confirmed by our optimized $P_{A}(n)$.


\begin{thebibliography}{99}
\bibitem{Bennett}
C. H. Bennett, G. Brassard, C. Crepeau, R. Jozsa, A. Peres, and W. K. Wootters., Phys. Rev. Lett. {\bf 70}, 1895 (1993).
\bibitem{Susskind}
L. Susskind and Y. Zhao, Phys. Rev. D {\bf 98}, 046016 (2018).
\bibitem{Schuster}
T. Schuster, B. Kobrin, P. Gao, I. Cong, E. T. Khabiboulline, N. M. Linke, M. D. Lukin, C. Monroe, B. Yoshida, and N. Y. Yao., Phys. Rev. X {\bf 12}, 031013 (2022).
\bibitem{Jafferis}
D. Jafferis, A. Zlokapa. J. D. Lykken, D. K. Kolchmeyer, S. I. Davis, N. Lauk, H. Neven, M. Spiropulu., Nature {\bf 612}, 51 (2022).
\bibitem{Popov}
A. Milekhin and F. K. Popov, arXiv:2210.03083.
\bibitem{Kitaev}
A. Kitaev, Ann. Phys. (NY) {\bf 321}, 2 (2006).
\bibitem{Nussinov}
Z. Nussinov and J. van den Brink, Rev. Mod. Phys. {\bf 87}, 1 (2015).
\bibitem{Hermanns}
M. Hermanns, I. Kimchi, and J. Knolle, Annu. Rev. Condens. Matter Phys. {\bf 9}, 17 (2018).
\bibitem{Knolle}
J. Knolle and R. Moessner, Annu. Rev. Condens. Matter Phys. {\bf 10}, 451 (2019).
\bibitem{Takagi}
H. Takagi, T. Takayama, G. Jackeli, G. Khaliullin, and S. E. Nagler, Nat. Rev. Phys. {\bf 1}, 264 (2019).
\bibitem{Janssen}
L. Janssen and M. Vojta. J. Phys.: Condens. Matter {\bf 31}, 423002 (2019).
\bibitem{Motome}
Y. Motome and J. Nasu, J. Phys. Soc. Jpn. {\bf 89}, 012002 (2020).
\bibitem{Hickey}
C. Hickey and S. Trebst, Phys. Rep. {\bf 950}, 1 (2022).
\bibitem{Minakawa}
T. Minakawa, Y. Murakami, A. Koga, and J. Nasu, Phys. Rev. Lett. {\bf 125}, 047204 (2020).
\bibitem{Koga}
A. Koga, Y. Murakami, and J. Nasu, Phys. Rev. B {\bf 103}, 214421 (2021).
\bibitem{Taguchi}
H. Taguchi, Y. Murakami, A. Koga, and J. Nasu, Phys. Rev. B {\bf 104}, 125139 (2021).
\bibitem{Nasu}
J. Nasu, Y. Murakami, and A. Koga, Phys. Rev. B {\bf 106}, 024411 (2022).
\bibitem{Misawa}
T. Misawa, J. Nasu, and Y. Motome, Phys. Rev. B {\bf 108}, 115117 (2023).
\bibitem{Takahashi}
M. O. Takahashi, M. G. Yamada, M. Udagawa, T. Mizushima, S. Fujimoto., Phys. Rev. Lett. {\bf 131}, 236701 (2023). 
\bibitem{Vijay}
S. Vijay and L. Fu, Phys. Rev. B {\bf 94}, 235446 (2016).
\bibitem{Fu}
L. Fu, Phys. Rev. Lett. {\bf 104}, 056402 (2010).
\bibitem{Huang}
H.-L. Huang, M. Narozniak, F. Liang, Y. Zhao, A. D. Castellano, M. Gong, Y. Wu, S. Wang, J. Lin, Y. Xu, H. Deng, H. Rong, J. P. Dowling, C.-Z. Peng, T. Byrnes, X. Zhu, and J.-W. Pan, Phys. Rev. Lett. {\bf 126}. 090502 (2021).
\bibitem{Hotta}
M. Hotta, Phys. Lett. A {\bf 372}, 5671 (2008).
\bibitem{Hotta2}
M. Hotta, Phys. Rev. D {\bf 78}, 045006 (2008).
\bibitem{Hotta3}
M. Hotta, J. Phys. Soc. Jpn. {\bf 78}, 034001 (2009).
\bibitem{Hotta4}
M. Hotta, Phys. Lett. A {\bf 374}, 3416 (2010).
\bibitem{Trevison}
J. Trevison and M. Hotta, J. Phys. A: Math. Theor. {\bf 48}, 175302 (2015).
\bibitem{Rodriguez}
N. A. Rodriguez-Briones, H. Katiyar, E. Martin-Martinez, and R. Laflamme., Phys. Rev. Lett. {\bf 130}, 110801 (2023).
\bibitem{Ikeda}
K. Ikeda, Phys. Rev. Appl. {\bf 20}, 024051 (2023).
\bibitem{Ikeda2}
K. Ikeda, AVS Quantum Sci. {\bf 5}, 035002 (2023).
\bibitem{Itoh}
K. Itoh, Y. Masaki, and H. Matsueda, arXiv:2305.03967.
\bibitem{Wang}
J. Wang and S. Yao, arXiv:2405.13886.
\bibitem{Sagawa}
T. Sagawa and M. Ueda, Phys. Rev. Lett. {\bf 100}, 080403 (2008).
\bibitem{Tajima}
H. Tajima, Phys. Rev. E {\bf 88}, 042143 (2013).
\bibitem{Funo}
K. Funo, Y. Watanabe, and M. Ueda, Phys. Rev. A {\bf 88}, 052319 (2013).
\bibitem{Park}
J. J. Park, K.-H. Kim, T. Sagawa, and S. W. Kim, Phys. Rev. Lett. {\bf 111}, 230402 (2013).
\bibitem{Manzano}
G. Manzano, F. Plastina, and R. Zambrini, Phys. Rev. Lett. {\bf 121}, 120602 (2018).
\bibitem{Minagawa}
S. Minagawa, M. H. Mohammadt, K. Sakai, K. Kato, and F. Buscemi, arXiv:2308.15558.
\bibitem{Itoh2}
K. Itoh, Y. Masaki, and H. Matsueda, arXiv:2408.11522.
\bibitem{Groenewold}
H. J. Groenewold, Int. J. Theor. Phys. {\bf 4}, 327 (1971).
\bibitem{Ozawa}
M. Ozawa, J. Math. Phys. {\bf 27}, 759 (1986).
\end{thebibliography}
\end{document}